\documentclass{emulateapj}
\newcommand{\etal}{{et al.~}}

\begin{document}

%\pagenumbering{arabic}
\title{Identification of the Microlens in Event MACHO-LMC-20}
\shorttitle{The MACHO-LMC-20 Lens}
\shortauthors{Kallivayalil \etal}

\author{Nitya Kallivayalil\altaffilmark{1}, Brian
  M.\ Patten\altaffilmark{1}, Massimo Marengo\altaffilmark{1}, Charles
  Alcock\altaffilmark{1}, Michael W.\ Werner\altaffilmark{2} and
  Giovanni G.\ Fazio\altaffilmark{1}}
\altaffiltext{1}{Harvard-Smithsonian Center for Astrophysics, 
60 Garden Street, Cambridge, MA 02138}
\altaffiltext{2}{Jet Propulsion Laboratory, California Institute of
  Technology, 4800 Oak Grove Drive, Pasadena, CA 91107}
\email{nkalliva@cfa.harvard.edu}

\begin{abstract}
We report on the identification of the lens responsible for
microlensing event MACHO-LMC-20. As part of a \textit{Spitzer}/IRAC
program conducting mid-infrared follow-up of the MACHO Large
Magellanic Cloud microlensing fields, we discovered a significant flux
excess at the position of the source star for this event. These data,
in combination with high resolution near-infrared
\textit{Magellan}/PANIC data has allowed us to classify the lens as an
early M dwarf in the thick disk of the Milky Way, at a distance of
$\sim 2$ kpc. This is only the second microlens to have been
identified, the first also being a M dwarf star in the disk. Together,
these two events are still consistent with the expected frequency of
nearby stars in the Milky Way thin and thick disks acting as lenses.
\end{abstract}
\keywords{Galaxy: structure --- gravitational lensing --- stars: late-type}

\section{Introduction}

We have been conducting near and mid-infrared follow-up of the MACHO (Massive
Compact Halo Object) Large Magellanic Cloud (LMC) microlensing event
fields to try to identify the lensing population. Identification of
the lenses can resolve questions regarding their locations, and can
support a microlensing interpretation of the events. Deep photometry
using \textit{Spitzer's} Infrared Array Camera (IRAC) of all the
lensed LMC stars reported by MACHO can be used to test the hypothesis
that the Milky Way has a thin or thick disk population of cool stars
that are responsible for the lensing.

Such a population would be readily detectable by \textit{Spitzer}, but
because IRAC has a spatial resolution of $\sim1.7\arcsec$ (FWHM), not
enough time has past since the original microlensing event to allow
the source and lens stars to separate via proper motions. However, the
presence of a cool lens can be inferred by way of an infrared (IR)
excess (Von Hippel \etal 2003). While this technique requires a number
of assumptions to be made about the lens and the source star, it has
had some success to date. In the case of MACHO-LMC-5 (hereafter Event
5), the lens properties inferred from the flux excess technique using
IRAC (Nguyen \etal 2004) were strikingly consistent with those
obtained by direct high-resolution \textit{HST} imaging (Alcock \etal
2001b; Gould 2004; Drake \etal 2004). The Event 5 lens was found to be
a M5 dwarf in the disk of the Milky Way, and was the only microlens to
have been identified to date. On its own it did not shed any light on
the microlensing population in general.

%MACHO-LMC-5 (hereafter Event 5) the lens is most likely a disk M5
%dwarf with a mass of $\sim0.2M_{\odot}$, as inferred from its IRAC
%fluxes and colors (Nguyen \etal 2004). Alcock \etal (2001a) used
%direct imaging to examine Event 5 using \textit{HST}/WFPC2 just 6
%years after the microlensing event, yielding a clear separation of the
%lens and source ($0.134\arcsec$). Analysis of these \textit{HST} data
%by Alcock \etal (2001b) and Gould (2004) along with similar data
%obtained with the \textit{HST}/ACS (Drake \etal 2004), yielded
%fundamental properties of the lens star that are strikingly consistent
%with the flux-excess technique results determined using
%\textit{Spitzer}/IRAC. This is the only microlens that has been
%identified to date. It is consistent with the number of events that
%was expected from the Milky Way disk, and thus on its own does not shed any
%light on the microlensing population in general nor does it signify
%any anomolous structure in the Milky Way disk.

In this letter we report the identification of one more lens from the
MACHO sample, that for MACHO-LMC-20 (hereafter ``Event 20''; Alcock
\etal 2000). Event 20 was a low-magnification event ($A_{\rm
max}=2.95$) that occurred in November, 1997. It only passed the looser
`criteria B' MACHO cuts which allowed events of lower signal-to-noise
than the stricter `criteria A' ones (Alcock \etal 2000). The event
duration was 72.7 days, which for a typical halo model, places the
lens mass at approximately $0.5M_\sun$ (Alcock \etal 1993). The
baseline (unmagnified) $V$ magnitude of the source$+$lens system was
21.35, with $V-R=0.57$. IRAC detected a clear IR excess at the source
location in both the 3.6 and 4.5$\micron$ bands relative to the
expected flux in these bands for a star of the above-mentioned
magnitude and color ($\sim0.6$ mag of excess flux), suggesting
blending with a foreground cool star. This data, in combination with
subsequent ground-based follow-up using the PANIC infrared imager on
Magellan, has enabled us to infer that the lens is an early M dwarf
approximately $2$ kpc away in the Milky Way thick disk. Together with
Event 5, the two events are consistent with the number of events
expected from the Milky Way disk, and thus do not signify any
anomalous structure in the disk. \S~2 briefly describes the IRAC data,
\S~3 describes the PANIC observations and analysis, and \S~4 presents
the results and a discussion of them.

\section{Description of Spitzer data in Brief}
MACHO-LMC-20 (hereafter Event 20) was observed using
\textit{Spitzer}/IRAC as part of PID 20121, an ongoing near-IR
follow-up investigation of the MACHO LMC fields (follow-up targets are
MACHO-LMC-1, 4, 5, 6, 7, 8, 9, 13, 14, 15, 18, 20, 21, 25 \& 27). IRAC
is a four-channel camera consisting of two pairs of $256\times256$
pixel InSb and Si:As IBC detectors that provide simultaneous images at
3.6, 4.5, 5.8, and 8$\micron$\footnote{IRAC 3.6 and 4.5$\mu$m bands
have effective wavelengths similar to the widely used L and M filters
respectively (Fazio \etal 2004).}. Each channel has a field of view of
$\sim5.12\arcmin\times5.12\arcmin$ and an $\sim 1.2\arcsec$ pixel
resolution. As mentioned in the introduction, at this plate-scale, the
source and lens star cannot be resolved even for lenses located
relatively nearby in the Galactic disk. However, a cool lens would
contribute a significant fraction of the source$+$lens (hereafter
designated `system') emission in the IRAC bandpasses. Note that while
we do have to consider how the microlensing system has evolved in time
since the peak magnification, the chance placement of the source star
near a foreground red object is very low (see e.g. Alcock \etal
2001b). Thus if the infrared excess coincides with the position of the
source star, we are very likely seeing the contribution from a cool
lens.

Further results on the IR follow-up will be published in
forthcoming papers (see Kallivayalil \etal 2004 \& Patten \etal 2005
for a summary of the progress so far). Here we focus on Event 20.  The
IRAC observations were centered on the event position. The target area
was imaged using a 12-position Reuleaux triangle dither pattern with 4
repeat exposures of 30 second FRAMETIME at each position. This
strategy produced background-limited images for each individual
exposure while the combination of repeats and dithers minimized the
impact of cosmic rays, bad pixels and other fixed-pattern noise in the
arrays. The total effective exposure time was 1440 seconds in each
IRAC band.

The data were reduced using the IRAC post-BCD (basic calibrated data)
processing software ``IRACproc'' (Schuster \etal 2006). In brief, the
software mosaics the basic calibrated data delivered by the Spitzer
Science Center and performs cosmic ray and bad pixel rejection. The
individual frames were visually inspected to ensure that the target
area was clean of cosmic rays and other blatant artifacts before
processing.

In order to identify the position of the source stars in our IRAC
data, we registered the cleaned mosaics with the MACHO $R$-band
discovery images\footnote{The finder charts for the MACHO LMC events
are available at http://wwwmacho.mcmaster.ca/.}. This was quite
straightforward as the MACHO and IRAC pixel scales are
comparable. Figure~1 shows the MACHO image of Event 20 on the top
panel. The crosshairs are centered on the system. For comparison, the
final mosaic of the IRAC 3.6$\micron$ data is shown on the bottom with
a circle highlighting the system. The IR excess is visible by eye. The
APPHOT package in IRAF was used to perform aperture photometry on the
cleaned mosaics, and aperture corrections were performed for the
source apertures that we employed. Table~1 shows the IRAC 3.6 and
4.5$\micron$ (hereafter $[3.6]$ \& $[4.5]$) photometry for the system
(which is dominated by the flux of the lens candidate). We had no
detections in the two longer IRAC bands.

\begin{figure}
\centering{
\includegraphics[angle=0,scale=0.42]{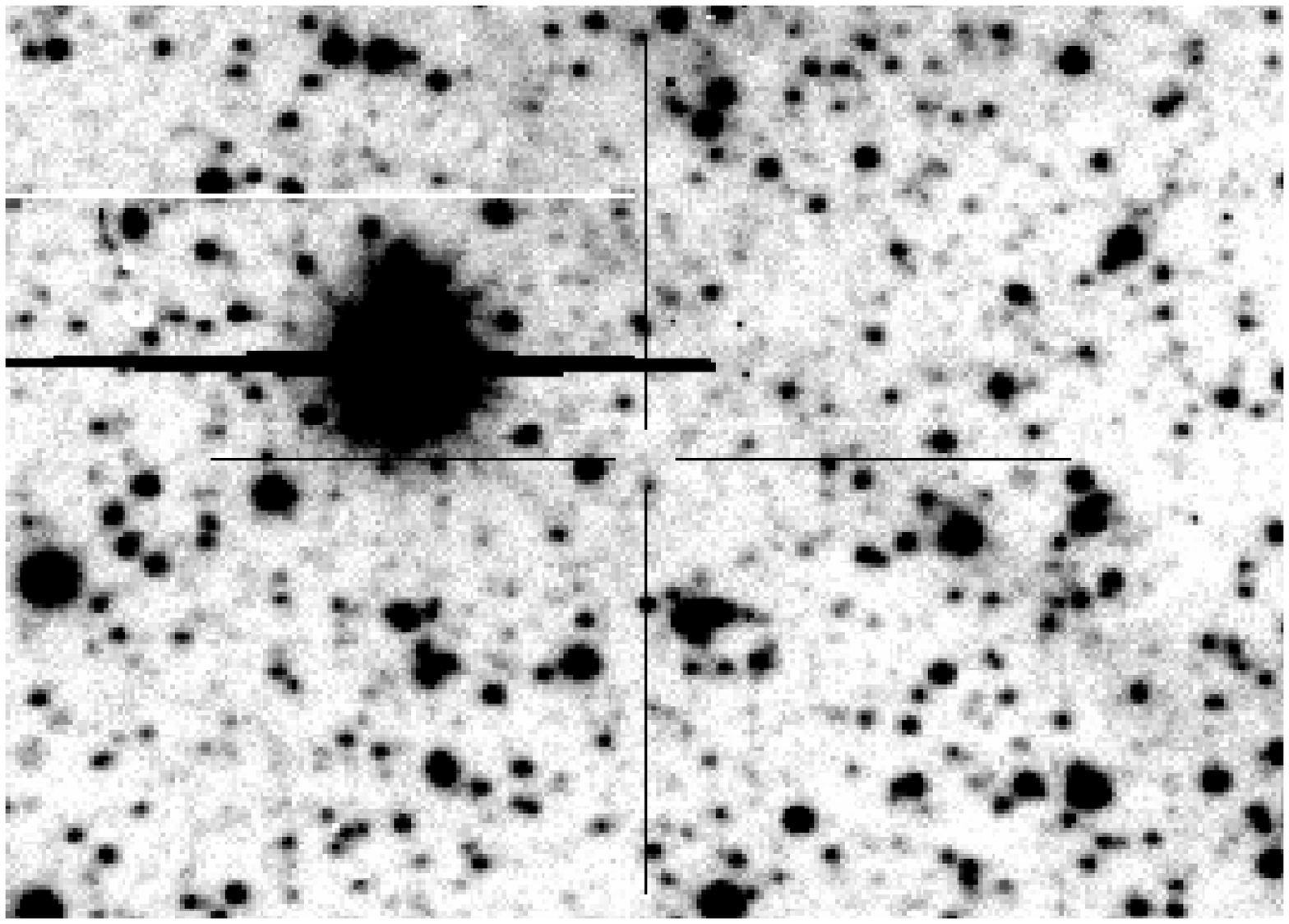}
\includegraphics[angle=0,scale=0.42]{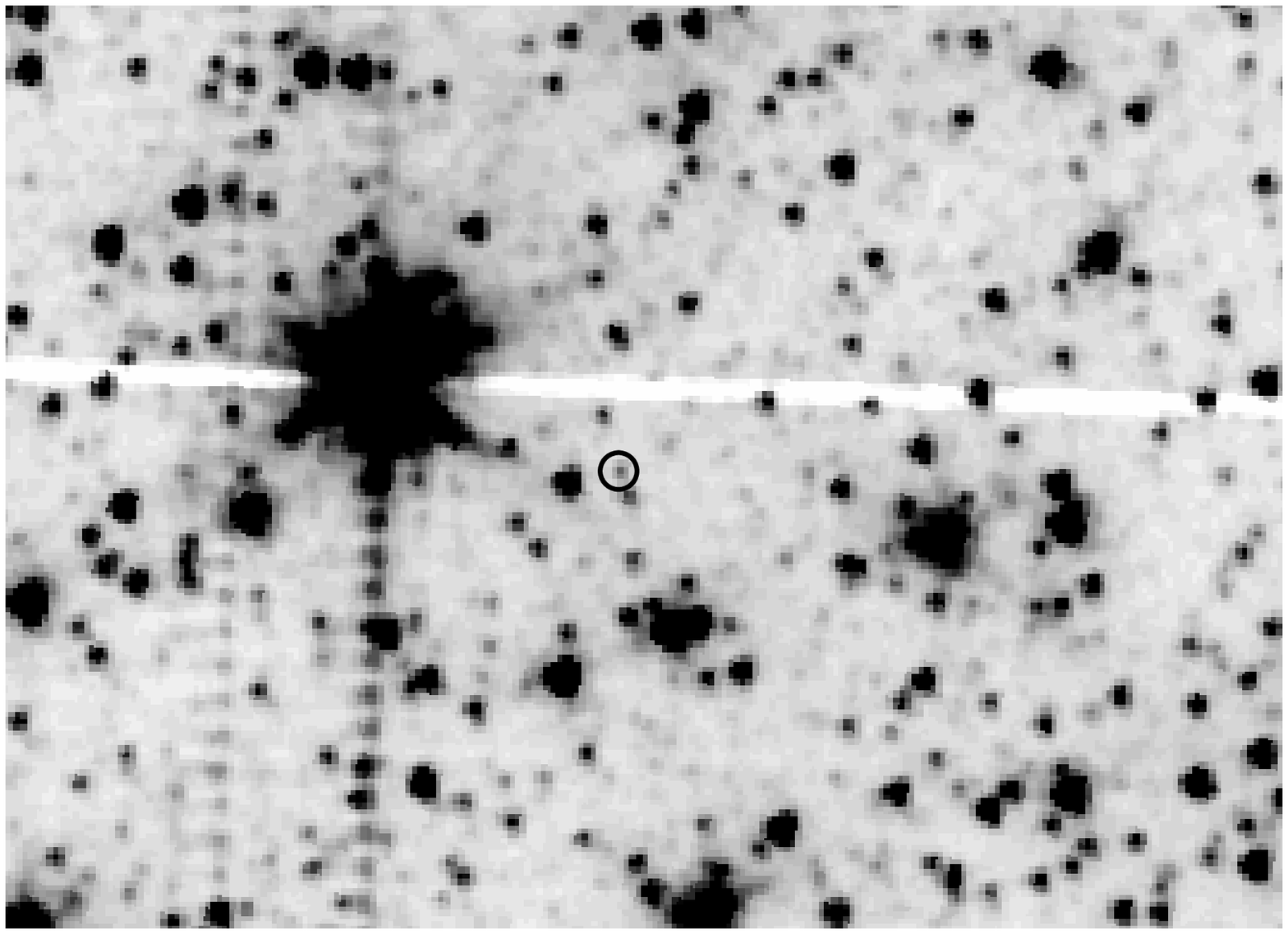}}
\caption{(\textit{Top}) The (baseline) $R$-band MACHO image of Event 20. The
  crosshairs are centered on the system. (\textit{Bottom}) The
  IRAC [3.6] image of the same field. The circle highlights the
  system.}
\end{figure}

\section{Description of PANIC Observations and Analysis}
With detections only in the first two IRAC bands, we were left with a
certain amount of color-degeneracy in determining whether the lens is
a M or L-dwarf, or even something more massive (see Patten \etal
2006). To address this problem we secured \textit{Magellan}/PANIC
(Persson's Auxiliary Nasmyth Infrared Camera) high resolution $J$, $H$
and $K_s$-band imaging in order to characterize the lens. The $JHK_s$
data in combination with IRAC would allow for a clearer discrimination
between M and L dwarfs, and the high resolution (0.125$\arcsec$ per
pixel, $128\arcsec \times 128\arcsec$ field-of-view) would allow us to
ascertain whether the IR excess was from a point-source like system,
as we expected, or from an unrelated neighbor.

We observed the field of Event 20 for most of the night on January 10,
2006. The source star is faint (see Table~1) and we had to integrate
on source for 45 minutes (60 second frames) in $J$-band, 50 minutes
(15 second frames) in $H$-band and 50 minutes (15 second frames) in
$K_s$-band in order to achieve S/N$>10$ in all bands. The use of a
5$\arcsec\times 5\arcsec$ step dither also allowed us to observe the
target star and reference stars at different places in the pixel
grid. The viewing conditions were non-photometric, with the seeing
changing from $0.7\arcsec$ to $0.4\arcsec$ and back up to $0.5\arcsec$
over the course of the night. We took repeated observations of 9
bright 2MASS (Skrutskie \etal 2006) comparison stars distributed
around the Event 20 field in order to get a good handle on how the sky
was varying over the night, and thus our systematic errors, and also
to tie all the photometry down to the 2MASS system. The 9 2MASS stars
were distributed in 3 fields, which we call fields A, B \& C for
future reference.

The data were reduced using the GOPANIC routine in the PANIC IRAF
software (Martini \etal 2004). The GOPANIC task provides a complete
pipeline for PANIC data, including summing the images in a loop,
applying a linearity correction, flat-fielding, sky subtraction and
distortion correction. The final images showed that the system is
still a point source, with no evidence of a displacement of the IR
source from the MACHO position. The DAOPHOT package in IRAF (Stetson
1987) was used to fit an empirical point spread function (PSF) to the
final images and to do photometry.

Our analysis strategy was simply to calculate a magnitude offset
(essentially a zero point) between the magnitudes that we measured for
the comparison stars in each of our fields and the known 2MASS
magnitudes, and then to use this offset in order to calibrate the
photometry of the lens candidate. One component of the final
photometric error would thus be the RMS of these field offsets, since
in photometric conditions, the offsets should all be the same (modulo
any variability in the 2MASS stars). Since conditions were not
photometric however, in addition to the zero-point offset we also
needed to calculate an additional field-dependent offset, which would
tie the 3 standard fields and the Event 20 field together. This was
done using the common stars in the fields. Fields A \& B had regions
of overlap with the Event 20 field and could thus be directly tied to
it. Field C overlapped with field B, and was tied to the field of
Event 20 via field B. Our final offset per field, $\Delta m$(field),
was then calculated from the 2MASS offset, $\Delta m_{\rm 2MASS}$, and
the field-dependent calibration term, $calib(\rm field)$, simply as
follows:
\begin{equation}
\Delta m {\rm(field)} = \Delta m_{\rm 2MASS} + calib(\rm field).  
\end{equation}

The $calib(\rm field)$ terms were typically small, consistent with
photometric variations caused by changes in the seeing.  They were
estimated from roughly 10 common stars in each overlap region. Once we
had an estimate of $\Delta m {\rm(field)}$ from fields A, B \& C, we
calculated a mean value that could be applied to our photometry for
the lens candidate. The RMS of $\Delta m {\rm(field)}$ is a direct
measure of both random and systematic photometric errors. We
calculated our final photometric error for each filter as the
quadrature sum of the photometric error of the lens candidate, and the
RMS of the $\Delta m {\rm(field)}$s. The typical final error is
$\sim0.05$ mag. Given these errors we did not think it necessary to
implement any higher order corrections to our calibration, such as
color terms. The PANIC photometry of the lens is presented in Table~1.

%%TABLE HERE%%%
\begin{deluxetable*}{ll}
\tabletypesize{\footnotesize}
%\rotate
\tablewidth{0pt}
\tablecolumns{2}
\tablecaption{Photometry for Event 20}
\tablehead{
\colhead{}  & \colhead{}}
\startdata
MACHO ID & 17.2221.1574\\
RA, DEC (2000) & 04 54 19, -70 02 15\\ 
$V$ & 21.35\\ 
$J$ & $18.84 \pm0.04$\\ 
$H$ & $18.28\pm0.06$ \\
$K_s$ & $18.06\pm 0.05$\\ 
$[3.6]$ & $18.02 \pm0.11$\\
$[4.5]$ & $18.37\pm 0.21$\\
\enddata
\tablecomments{The data for the first three rows are from Alcock \etal 2000.}
\end{deluxetable*}

\section{Results \& Discussion}
The suggestion of microlensing by the Galactic dark halo (Paczynski
1986) was followed-up by many teams. While the MACHO collaboration
reported 16 microlensing events towards the LMC (Alcock \etal 1993,
1997, 2000), the EROS collaboration has found only 3 (Aubourg \etal
1993; Lasserre \etal 2000). The MACHO efficiency analysis indicates a
dark halo with a MACHO fraction of 20\% which corresponds to an
optical depth of $\tau=1.2^{+0.4}_{-0.3} \times 10^{-7}$ due to lenses
of $\sim0.5M_{\odot}$. This rate is \textit{significantly} higher than
what was expected from known Galactic and LMC stellar populations; the
latter have $\tau$ between $0.24 \times 10^{-7}$ and $0.36 \times
10^{-7}$ (Alcock \etal 2000; Griest \& Thomas 2005). The upper limit
on the EROS microlensing optical depth is only barely consistent with
that of MACHO. Recently, the EROS-2 project reported their results,
and combined with the results of EROS-1, they report an optical depth
toward the LMC of $\tau<0.36\times10^{-7}$ (95\% confidence) due to
lenses of $\sim0.4M_{\odot}$, corresponding to a halo mass fraction of
less than 7\% (Tisserand \etal 2006).

% There has been much progress since gravitational
%microlensing was first clearly detected towards the Large and Small
%Magellanic Clouds (LMC \& SMC) (Afonso \etal 2003; Alcock \etal 1993,
%1997, 2000; Lasserre \etal 2000; Aubourg \etal 1993; Udalski \etal
%1993) but the interpretation of these events is still shrouded in
%mystery. The Magellanic Cloud events probe the contribution of MACHOs
%(Massive Compact Halo Objects) to the dark matter in the halo of the
%Milky Way (Paczynski 1986). The MACHO collaboration reported 16 bona

%The EROS collaboration (Aubourg \etal 1993; Lasserre \etal 2000) which
%also monitored the LMC has found only 3 microlensing events, and their
%upper limit on the microlensing optical depth is only barely
%consistent with that of the MACHO collaboration. 

%which used a
%subsample of $7\times10^6$ bright stars from their much larger sample
%in order to minimize backgrounds due to variable stars, reported their
\begin{figure}
\begin{center}
\includegraphics[angle=0,scale=0.42]{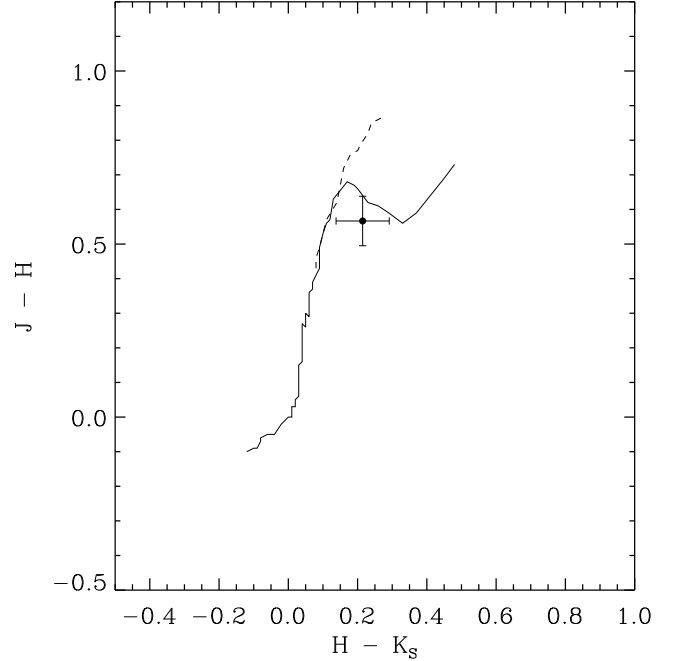}
\caption{A plot in $J-H$ vs.~$H-K_s$--space showing the intrinsic
  relation for dwarfs (solid line) and giants (dashed line). The
  filled circle with error bars shows our PANIC photometry for the
  Event 20 system. }
\end{center}
\end{figure}

Several groups have investigated the excess of microlensing towards
the LMC seen by MACHO, attributing it to as yet undetected structure
in the Milky Way (Rahvar 2005; Nguyen \etal 2004; Gates \& Gyuk 2001;
Evans \etal 1998; Zhao 1998; Zaritsky \& Lin 1997) or in the LMC
itself (Wu 1994, Sahu 1994; Alves \& Nelson 2000; Evans \& Kerins
2000; Di Stefano 2000; Gyuk \etal 2000; Zhao \etal 2003; Mancini \etal
2004; Alves 2004; Nikolaev \etal 2004). However, over and above the
question of the population responsible for the lensing, the
discrepancy between the MACHO and EROS results have also cast doubt
on whether the MACHO events are microlensing at all, or just
contamination from variable stars and background supernovae. Belokurov
\etal (2003, 2004) applied a neural network analysis to the MACHO data
in an attempt to develop independent selection criteria for
microlensing. They selected only 10 of the candidates that MACHO
picked out as bona fide microlensing. Their verdict on the
microlensing events as well as the original MACHO verdict are
presented in Table~4 of Bennett \etal (2005).

%%%

\begin{figure*}
\begin{center}
\includegraphics[angle=0,scale=0.65]{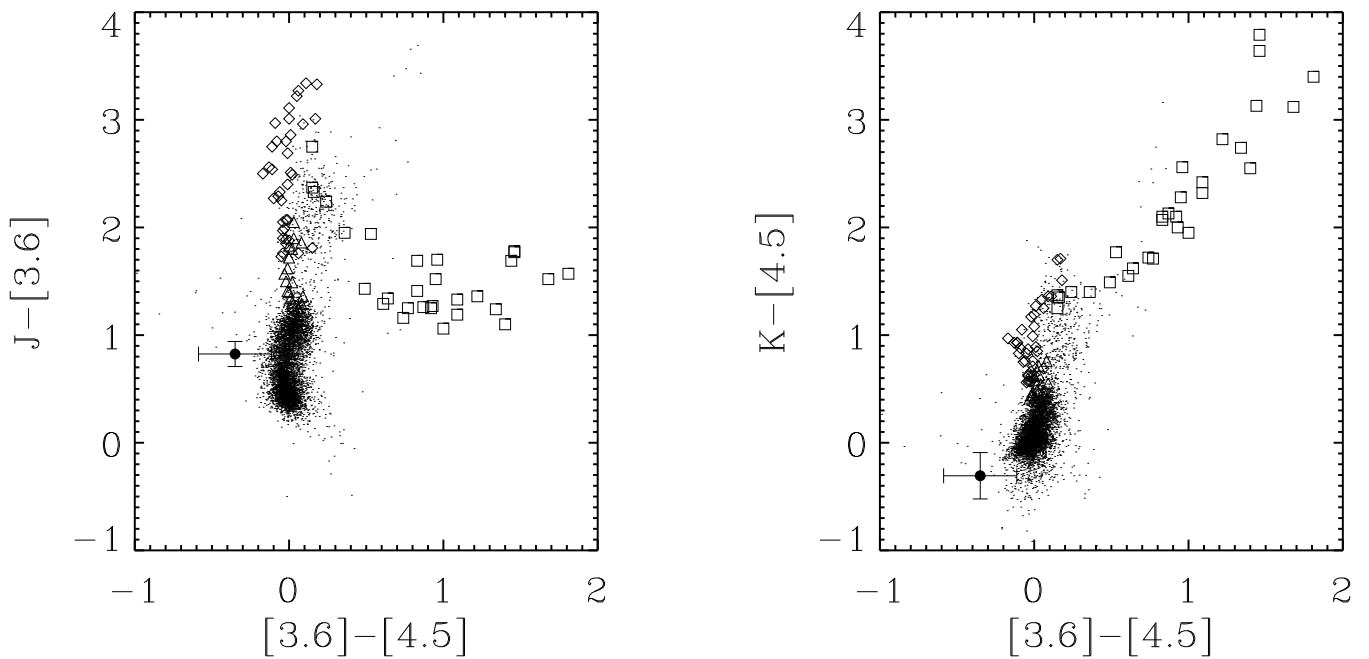}
\caption{(\textit{Left}) A plot showing $J-[3.6]$ vs.~$[3.6]-[4.5]$ for
  stars and galaxies in the First Look Survey. Late-type M dwarfs from
  the Patten \etal (2006) sample are overlaid using triangles, L
  dwarfs using diamonds and T dwarfs using squares. The filled circle
  with error bars shows are PANIC$+$IRAC photometry for the Event 20
  system. (\textit{Right}) The same objects are shown in $K-[4.5]$
vs.~$[3.6]-[4.5]$-space}
\end{center}
\end{figure*}

\begin{figure}
\begin{center}
\includegraphics[angle=0,scale=0.45]{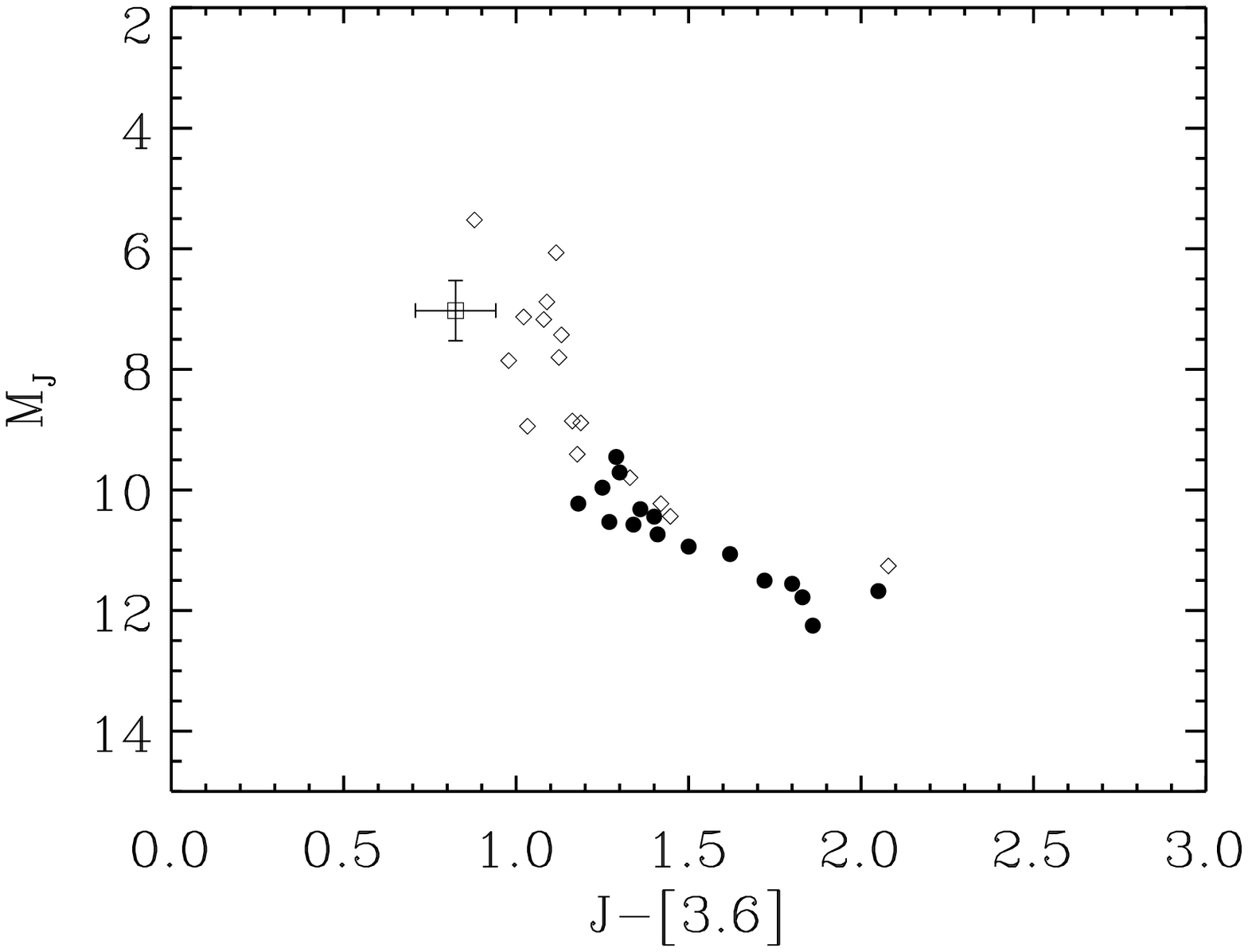}
\caption{A plot showing $M_J$ vs.~$J-[3.6]$ for
  M dwarfs (and 2 K7 dwarfs) from the Patten \etal (2006) sample
  (filled circles), and some earlier type M dwarfs from M. Schuster
  (private communication) (diamonds).  The square with error bars
  shows the lens candidate, which is consistent with where early M
  dwarfs lie on this relation. }
\end{center}
\end{figure}

In the case of the event at hand, Belokurov \etal (2003) used their
neural network to classify it as a supernova (SN) rather than
microlensing. However, Bennett \etal (2005) pointed out that Event 20
is a good disk lens candidate because Figure~7 of Alcock \etal (2000)
shows that, like Event 5, Event 20 is quite red for its brightness,
i.e. it appears to be blended with a star that is much redder than
other LMC stars of similar magnitude.

Our data allows us to weigh into this debate. Figure~2 is a $J-H$
vs.~$H-K_s$ plot, showing the intrinsic relation for dwarfs (solid
line) and giants (dashed line; dwarf data from Leggett 1992 and Kenyon
\& Hartmann 1995; giant data from Bessell \& Brett 1988) . The filled
circle with error bars shows our PANIC result for the lens, which
appears to be consistent with an early-M dwarf star. The mean
extinction at the location of Event 20 in the LMC is $A_V \sim 0.5$
mag (Zaritsky \etal 2004). This implies that, even if our M dwarf is
at the distance of the LMC (in which case we certainly would not see
it with either PANIC or IRAC), the mean extinction in $K$-band would
be $A_K \sim 0.05$. This is within our photometric errors, and thus we
do not account for reddening. Note, however, that in this and all
following analysis plots, the lens object is still blended with the
LMC source star. At these infrared wavelengths we expect that the
contamination from the source flux is minimal ($\sim 10\%$). The data
thus seem to corroborate Bennett's prediction of this event has
lensing by a low-mass star.

However, as suggested by Belokurov \etal (2003), we also consider the
possibility that the magnification in the MACHO light curve was caused
by a background SN. If this were the case, at $\sim9$ years since the
event we would now be seeing the host galaxy. In Figure~3 we address
this by showing the near-IR color-color space of objects in the
\textit{Spitzer}/IRAC First Look Survey (Lacy \etal 2005) for which
2MASS $J$ and $K$ photometry is available. The panel on the left shows
$J-[3.6]$ vs.~$[3.6]-[4.5]$. The points are stars and galaxies, with
the blue end of the plot comprising mostly main sequence stars. The
galaxies tend to be redder and extend into the diffuse plume at the
red end of the $y$-axis. Overlaid are late-M dwarfs (spectral type M6
and later) (triangles), L dwarfs (diamonds) and T-dwarfs (squares)
from Patten \etal (2006). The filled circle with error bars shows the
IRAC$+$ PANIC photometry for the lens. The lens lies in a part of the
color-color space that is more consistent with stars than with
galaxies, and within the large errors propagated through from the IRAC
photometry, it is consistent with an early M dwarf. The panel on the
right shows $K-[4.5]$ vs.~$[3.6]-[4.5]$. Again, the points at the blue
end of the plot are mostly stars with galaxies extending into a red
tail. In this color-space as well the data suggest that the lens is a
M dwarf and not a background galaxy. The placement of the lens could
be consistent with a blue galaxy, however, at the $JHK_s$ magnitude of
the lens, we would expect that galaxies are discernible from stars in
the high resolution PANIC images.

Thus our follow-up data confirm not only the microlensing nature of
this event but also that it was caused by a low-mass star. Finally,
Mancini \etal (2004) have made predictions, based on detailed models
of the LMC, about which of the MACHO events were likely due to lenses
in the LMC itself. They picked events 6, 8, 13 \& 14 as most likely
due to self-lensing because of their locations in the LMC. Event 20
was not picked as a high probability self-lensing candidate. The
Mancini \etal (2004) predictions are tabulated and compared against
the MACHO and Belokurov \etal (2004) predictions in Bennett \etal
(2005).

Given our measurement of the spectral type of the lens, we estimate
its distance from a sample of early M dwarfs using 2MASS $JHK_s$ and
[3.6] photometry provided by M. Schuster (private communication). If
the M dwarf is in the range M0--3.5, then its absolute magnitude would
be in the range $M_J = 7\pm0.5, M_H=7\pm0.5, M_{K_s}=6\pm0.4,$ and
$M_{[3.6]}=6\pm0.3$. In addition to this line of argument, in Figure~4
we show $M_J$ versus $J$-[3.6] for these M dwarfs (plus two K7 dwarfs
and additional late-type M dwarfs from M. Schuster) (triangles) and
for some late-type M dwarfs from Patten \etal (2006) (filled
circles). The square with error bars is the lens and it is clear that
its $J$-[3.6] color is consistent with the absolute magnitude expected
for an early M dwarf ($M_J\sim7$). The range of absolute magnitudes
quoted above imply a distance to the lens of $2\pm0.7$ kpc where the
error includes contributions from the range in spectral types and
the photometric error in the $J$-[3.6] color. The lens is thus in the
thick disk of the Milky Way\footnote{Events 20 \& 5 are evidence that
nearby dim stars can be studied through such monitoring programs
(``mesolensing'', Di Stefano 2005).}.  This is the second Milky Way disk lens
in the MACHO sample, the first being that of Event 5, which was closer
at $\sim600$ pc. A total of 0.75 events were expected from the Milky
Way thin and thick disks (Bennett \etal 2005), and thus these two disk
lenses are still consistent with the expected number. Event 20 did not
pass the MACHO criteria set `A' and so does not affect the conclusion
to the halo fraction made from it.

In a future paper we will present the full consensus of the near-IR
follow-up of the MACHO LMC microlensing fields. The main result of
this project thus far is that two of the microlensing events were
caused by foreground disk lenses. This is consistent with what was
expected from the Milky Way thin and thick disks.

%Bennett \etal (2005) state that the Poisson probability
%of
%detecting two or more disk events when only 0.75 were expected is
%17\%. Could this be evidence for a thicker thick disk than implied by
%current Milky Way models?

%Finally, we consider what this means for the MACHO halo fraction. 
%criteria set `B'. (XX need to talk to Charles, perhaps could help with
%this part of the conclusion.)
\acknowledgments The authors would like to thank M. Schuster for
providing IRAC photometry of some nearby M dwarfs. This work is based
[in part] on observations made with the Spitzer Space Telescope, which
is operated by the Jet Propulsion Laboratory, California Institute of
Technology under a contract with NASA. Support for this work was
provided by NASA through an award issued by JPL/Caltech. This
publication also makes use of data products from the Two Micron All
Sky Survey, which is a joint project of the University of
Massachusetts and the Infrared Processing and Analysis
Center/California Institute of Technology, funded by the National
Aeronautics and Space Administration and the National Science
Foundation, and the SIMBAD database, operated at CDS, Strasbourg,
France.

{\it Facilities:} \facility{Magellan:Baade (PANIC)}, \facility{Spitzer (IRAC)}

\end{document}